\documentclass[traditabstract]{aa} % for the abstract without structuration 
\usepackage{graphicx}

\usepackage{txfonts}
\usepackage{natbib}
\bibpunct{(}{)}{;}{a}{}{,} % to follow the A&A style

\begin{document}
   \title{Experiments on centimeter-sized dust aggregates and their implications for planetesimal formation}
	 \titlerunning{Experiments on centimeter-sized dust aggregates}
   \author{T. Meisner
          \inst{1} 
          \and
          G. Wurm
          \inst{1}
          \and
          J. Teiser\inst{1}
          }

   \institute{Faculty of physics, University of Duisburg-Essen,
              Lotharstr. 1, D-47057 Duisburg\\
              \email{thorsten.meisner@uni-due.de}
             }

   \date{Received ; accepted }
   
   %\abstract{}{}{}{}{} 
%5 {} token are mandatory
 
  \abstract{The first macroscopic bodies in protoplanetary disks are dust aggregates. We report on a number of experimental 
   studies with dust aggregates formed from micron-size quartz grains. 
We confirm in laboratory collision experiments an earlier finding that producing macroscopic bodies 
by the random impact of sub-mm aggregates results in a well-defined upper-filling factor of 0.31 $\pm$ 0.01. 
Compared to earlier experiments, we increase the projectile mass by about a factor of 100. The collision experiments also show
that a highly porous dust-aggregate can retain its highly porous core if collisions get more energetic and a denser shell forms
on top of the porous core. 
We measure the mechanical properties of cm-sized dust samples of different filling factors 
between 0.34 and 0.50. 
The tensile strength measured by a Brazilian test, varies between 1\,kPa and 6\,kPa.  
The sound speed is determined by a runtime measurement to range between 80 m/s and 140 m/s while Young's modulus is derived from the sound speed and 
varies between $7 $\,MPa and $25 $\,MPa. The samples were also subjected to quasi-static omni- and uni-directional compression
todetermine their compression strengths and flow functions. Applied to planet formation, our experiments  provide basic data for future simulations, explain the specific collisional outcomes observed in earlier experiments, and in general support a scenario where collisional growth of planetesimals is possible.}  

% context heading (optional)
  % {} leave it empty if necessary  
  % {}
  % aims heading (mandatory)
  % {...}
  % methods heading (mandatory)
  % {...}
  % results heading (mandatory)
  % {...}
  % conclusions heading (optional), leave it empty if necessary 
  % {}

   \keywords{Methods: laboratory -- Protoplanetary disks -- Planets and satellites: formation
   %giant planet formation --
                %$\kappa$-mechanism --
                %stability of gas spheres
               }

   \maketitle

\section{Introduction}

Collisions have an important influence on the evolution of planetary systems from their early beginnings onward. Understanding the effect of a collision requires knowledge about the mechanical properties of the colliding objects of given sizes. The different size ranges considered for the  
very early phases of planet formation are frequently termed pebbles, rocks, or boulders, though the mechanical properties of the initial population do not reflect these names. Initially, all objects can be considered as dust aggregates. As bodies smaller than planetesimal size ($\sim$ km-size) couple sufficiently well to the gas of a protoplanetary disk, collision velocities stay moderate and 
'only' reach several tens of m/s \citep{WeidenschillingCuzzi1993}. At these velocities, collisions are in general insufficiently energetic to melt or petrify an object. Therefore, originating from dust particles in a protoplanetary disk,  the initial population of objects have to be dust aggregates of a wide range of sizes. 

Two different approaches currently exist to simulate collisions between dust aggregates. The first method is based on molecular dynamics and simulates the interaction between each dust grain in two colliding aggregates. This method is restricted to small ($\ll 1$\,mm) dust aggregates owing to computational limits \citep{DominikTielens1997, WadaEtal2009, SeizingerEtal2012}. The other approach uses smoothed particle hydrodynamics (SPH) to simulate dust aggregates \citep{SchaeferEtal2007, BenzAsphaug1999, Sirono2004}. The single grains are no longer resolved in this method, but aggregates are characterized in terms of macroscopic mechanical properties, such as their compressive strength, tensile strength, or Young's modulus \citep{GeretshauserEtal2011}. The SPH-method is not restricted to a certain particle size, hence can provide crucial input for global coagulation models. However, the mechanical properties of protoplanetesimals are first needed as input for the simulations. One way to obtain these
parameters is to measure them in laboratory experiments.

So far, only a few studies of the mechanical properties of dust aggregates have been performed. 
\citet{BlumSchraepler2004} performed the first experiments to produce and characterize highly porous dust-aggregates (produced by random ballistic deposition) with a filling factor of 0.15 for a dust material consisting of monodisperse, spherical silica grains ($d = 1.5 \mu$m). They determined the evolution of the filling factor during an uni-directional compaction of the dust aggregates. In a later study, these experiments were extended to different dust materials \citep{BlumEtal2006}. In both of these studies three different dust samples were used (monodisperse silica spheres, diamond powder, and polydisperse quartz) and the porosity evolution at increasing uni-axial pressure was determined. Additionally, the tensile strength was measured for highly porous samples, produced by random ballistic deposition. \citet{GuettlerEtal2009} determined the omni-directional compression, but only for the monodisperse dust sample previously studied by \citet{BlumSchraepler2004} and \citet{BlumEtal2006}.  
The mechanical properties strongly depend on both the shape and size of the monomers. This was shown experimentally by \citet{BlumEtal2006} and theoretically by \citet{BertiniEtal2009}.  

The experiments presented here are dedicated to providing the mechanical properties for a polydisperse dust material with a grain size distribution, which might be typical of protoplanetary disks. Different experimental methods are applied to measure critical parameters such as tensile strength, sound velocity, or compressive strength. The mechanical properties are studied for various filling factors. 

Besides determining the basic properties of small dusty bodies, this work contains experimental results on porosities that might be expected for small sub-meter size protoplanetesimals. Further experiments also show that an evolving protoplanetesimal is unlikely to be homogeneous but we show that core - shell particles with highly porous cores and more compact shells can easily form through collisions with small mm-size aggregates.

The paper is organized as follows. We first briefly describe the dust samples and their preparation as generally as possible in section 2. We then give details of the experiments carried out with these samples in
section 3. Results and some general interpretation are given in section 4 and the final section 5 places the results and conclusions in the context of application in protoplanetary disks.

\section{\label{sec:sample_preparation}Sample preparation}

For all experiments, we used quartz dust with particle sizes ranging between 0.1 $\rm{\mu}$m and 10 $\rm{\mu}$m (80\% of the mass consisting of particles of size between 1\,$\rm{\mu}$m and 5\,$\rm{\mu}$m) from which we produced larger cm-sized aggregates. The same quartz dust had been used in several previous experiments \citep{BlumEtal2006, wurm2005, TeiserWurm2009a}.
We studied a number of different properties of the dust aggregates depending on the (volume) filling factor
\begin{equation}
\Phi = \frac{V_{solid}}{V_{total}} ,
\end{equation}
which is the ratio of the volume filled by solid material ($V_{solid}$) to the total volume ($V_{total}$). 
This is equivalent to the ratio of the density of the total object to the density of its solid quartz building-stones, which is $2.6\,\rm{g/cm^{3}}$. The samples for further measurements were always formed by compression of dust powder from all sides as the sample was confined to a cylindrical mold. Pressure was applied to the top face of the cylinder, which was free to move (Fig. \ref{fig:uniaxial}a). During the preparations, the applied omnidirectional pressure $\sigma_1$ was measured using a force sensor attached to the compressing piston (Fig. \ref{fig:uniaxial}a). After  correction for the exerted pressure to the additional weight of the top face bolt, $\sigma_1$ represents the compression stress that the dust aggregate was exposed to (Fig. \ref{fig:uniaxial}b). The maximum pressure applied was 55\,kPa.
 We prepared cylinders with constant masses of 20\,g that always had a diameter of 30\,mm. For this reason, their heights varied between 22\,mm and 36\,mm for different
 filling factors. After removing the dust cylinders from the mold, the volumes and detailed masses were measured. With this information, densities and therefore filling factors could be determined.  In total, the filling factors of the analyzed dust samples ranged from 0.24 to 0.51. The presented values for filling factors are mean values for whole samples. For the more porous samples in particular it is unclear \textit{a priori} whether the dust cylinders are homogeneous. To estimate any inhomogeneities, we produced dust cylinders with different filling factors in the range and way described above. We cut them into two halves with respect to 
 height and measured their filling factors in the same way. Each top half, which was more affected by the pressure applied to the top face of the cylinder was slightly more dense than the bottom half. The difference between the bottom and top halves was about 0.02 in filling factor, independently of the calculated mean value within our investigated range. These inhomogeneities -- which are $\pm 0.01$ deviations from the mean -- are small compared to the overall variations in the filling factor and are neglected in the following analysis.

\section{\label{sec:experimental_setup}Experimental setups}

The following different experiments were carried out to specify the elastic properties of the dust aggregates and some general trends in the evolution of the filling factor.
As far as the mechanical properties are concerned, we studied:
\begin{itemize}
\item omni- and unidirectional compression,
\item tensile strength,
\item sound velocity,
\item modulus of elasticity (Young's modulus) based on the sound velocity measurements.
\end{itemize}
Besides the determination of these basic properties, collision experiments were carried out to show the influence of collisions on the evolution of the filling factors of a forming dust aggregate.

\subsection{\label{sec:UCT}Procedure for compression measurements}

\begin{figure}[h]
\includegraphics[width=\columnwidth]{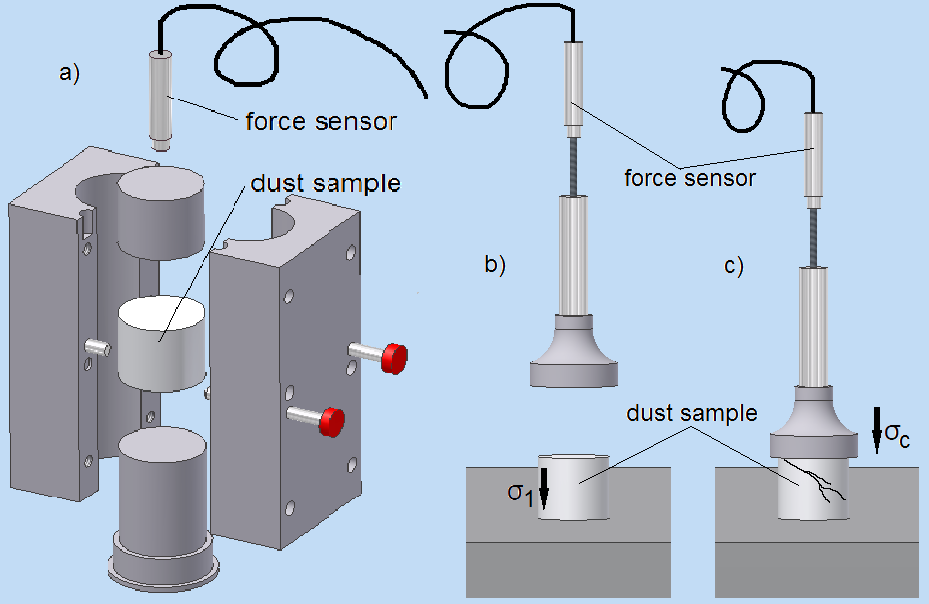}
    \caption{\label{fig:uniaxial} Experiment to measure the compression strength of a dust sample: \textbf{a)} production of a cylindrical dust- aggregate by applying a certain solidification pressure to a confined dust sample; \textbf{b)} removal of dust cylinder produced by a given compression stress $\sigma_{1}$; \textbf{c)} the compression strength $\sigma_{c}$ is measured when the sample breaks \citep{Jenike1964}.}
\end{figure}
Cylindrical dust-aggregates were produced as described in section \ref{sec:sample_preparation}. 
To these cylinders, we applied an unidirectional pressure at the top face and determined the threshold pressure $\sigma_c$ (equivalent to compression strength) when the sample did break as sketched in Fig. \ref{fig:uniaxial}c. We corrected the exerted pressure to the additional weight of the piston and obtained the uniaxial compression strength for each precompacted dust cylinder.
Samples were used with different precompactions i. e. filling factors ranging from 0.24 to 0.42.

\subsection{\label{sec:BT}Procedure for tensile strength measurements}
\begin{figure}[h]
\includegraphics[width=\columnwidth]{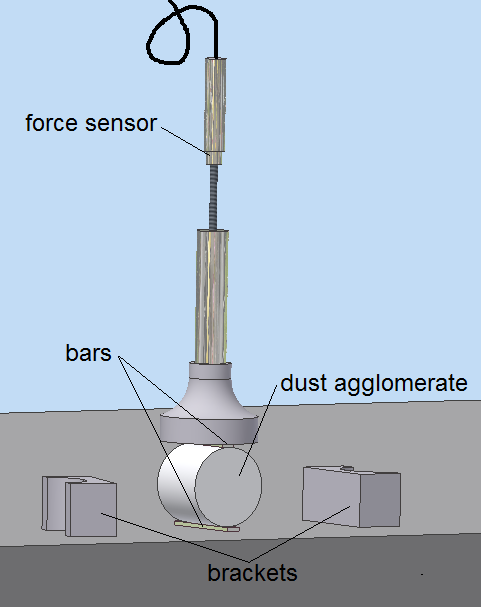}
    \caption{\label{fig:brazil} Setup for the Brazilian test: a piston connected with a force sensor is pushed down onto a bar lying above the sample. The two bars are used for a linear load distribution. The brackets hold the cylinder in place initially. Pressure is applied until the aggregate breaks.}
\end{figure}

To determine the tensile strength of the dust cylinders, we performed Brazilian tests. We note that independent but apparently similar measurements have been carried out by Kothe et al. (personal communication) for the same type of dust and it will be interesting to perform a
detailed comparison with their results in the near future.
The essence of the Brazilian test is that a pressure is imposed upon the lateral surface of the dust 
cylinders (Fig. \ref{fig:brazil}) in terms of two opposing effective single-loads. The pressure at which the sample breaks is then determined.  

\begin{figure}[h]
\includegraphics[width=\columnwidth]{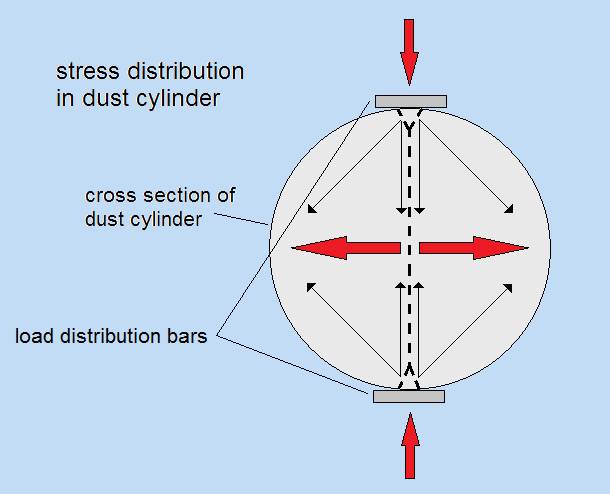}
    \caption{\label{fig:fracpat}Illustration of the load distribution in a dust cylinder during the performance of the Brazilian test. 
    The fracture pattern is given along the dashed lines according to \citet{Mitchell1961}.}
\end{figure}

A linear distribution of an applied load must be provided as illustrated in Fig. \ref{fig:fracpat}, to prevent too large compression stresses at the point of load application. It is necessary to produce a tensile fracture when the dust cylinder is crushed. According to \citet{Mitchell1961}, a fracture of a breaking dust cylinder in a splitting tensile test can take different shapes depending on the width of the load distribution bars. While the non-use of bars causes a stress fracture, only the use of narrow bars generates a vertical division line characteristic of a proper fracture, which is the indication of a tensile fracture. In practice, two small wedged pieces of dust are produced at the break. They develop directly above and below the bars and indicate a shear fracture. We used bars with a width of 5\,mm and a height of 2\,mm. Single vertical fracture lines were always produced from nearly the top bar to the bottom bar. The detailed width or contact area does not enter into the final determination of the tensile strength as only two times the applied force and the mantle surface of the dust cylinder enter according to

\begin{equation}
\rm{\sigma_{ts}} = \rm\frac{2 \cdot F}{\pi d L} ,
\end{equation}
for diameter $d$ and length $L$ of the cylinder.

The bars were placed between the dust cylinder used and the acting loads from the piston and the table. The dust aggregate was confined between two brackets to avoid rolling. On top, a piston and a force sensor were attached to one bar to determine the splitting force. The two brackets were removed as soon as the piston is in place and the load sufficient to keep the dust cylinder in place. Pressure was then applied. 
In addition to the pressure measurements, we used a high speed camera to observe the formation of the fracture and the break-up of the samples.    
For these experimental series, samples were used for a range of volume filling-factors from 0.37 to 0.51.

\subsection{\label{sec:MSV}Procedure for sound velocity measurements}          
\begin{figure}[h]
\includegraphics[width=\columnwidth]{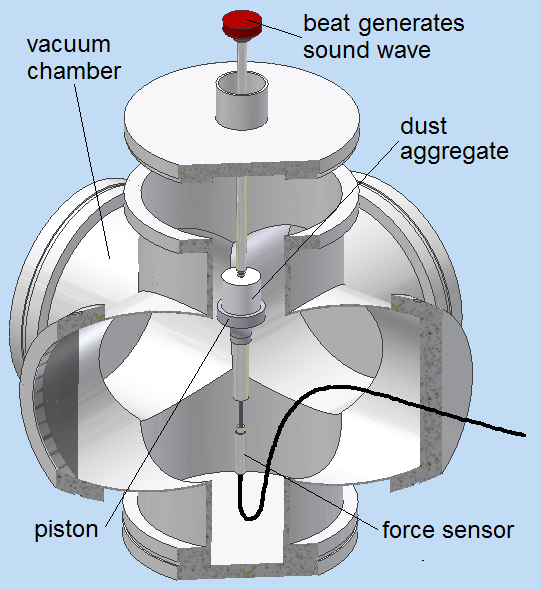}
    \caption{\label{fig:schall} Setup for measurement of sound velocities in dust samples inside a vacuum chamber}
\end{figure}

The sound velocity in the dust cylinders was measured in a basic way. A sound wave was initiated through a mechanical contact on one
side of the aggregate by manually knocking against
the aggregate support structure. Our results do not depend on any variations in the manual knocking.
As the sound velocity was known to be lower than the speed of sound in air, we measured the sound velocity in a vacuum chamber below
$4 \cdot 10^{-2}$\,mbar (Fig. \ref{fig:schall}).  
The time when the pulse is generated was recorded, the pulse itself closing an electrical contact. The arrival time of the signal at a force sensor on the other side of the aggregate was also recorded. The time delay was corrected for sound-wave travel times using the attached hardware. With the predetermined aggregate size, the sound velocity was calculated.  
We averaged 30 to 40 measurements for a given aggregate. These measurements were performed with dust cylinders that varied in terms of their filling factors across the range from 0.33 to 0.51.

\subsection{\label{sec:DTS}Procedure for collision experiments with dust particles}           

The previous measurements were restricted to determining the properties of dust aggregates that were  
artificially formed with a defined volume filling-factor resulting from the omnidirectional compression within the mold. 
With protoplanetary growth in mind, compression might however result from collisions. A likely impact in the early formation phases is the impact of a smaller body (projectile) with a larger body (target). This impact can be described by neither uni- nor omnidirectional static compression.
Building a dust aggregate from scratch by impacts of sub-mm dust aggregates, \citet{TeiserEtal2011b} found that homogeneous 
aggregates form with a volume filling-factor that only depends on the impact velocity and reaches a limiting filling-factor 
of 0.31 at an impact velocity of 6 m/s. We used the same setup here. 
\begin{figure}[h]
\includegraphics[width=\columnwidth]{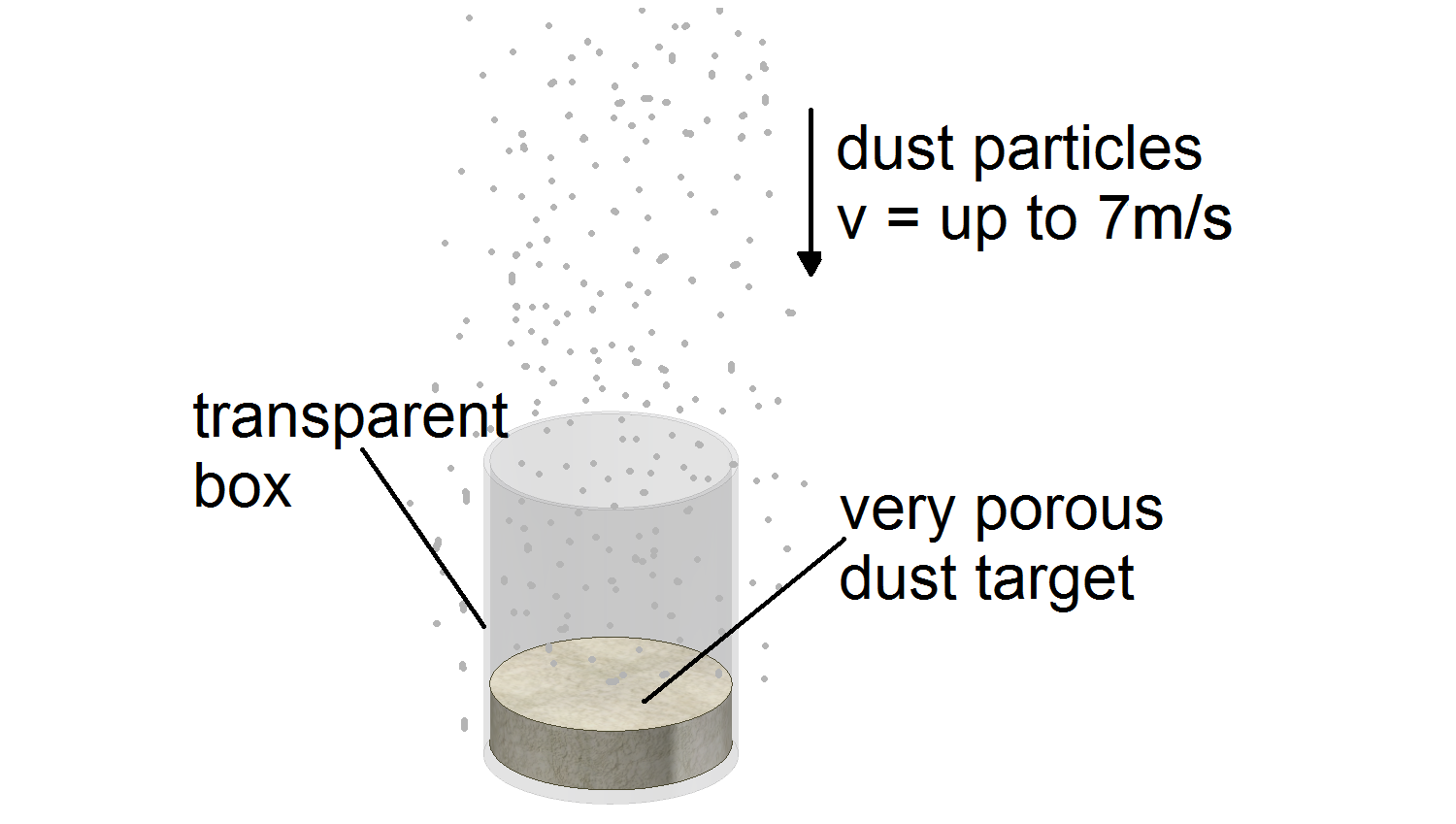}
    \caption{\label{fig:uneffecteddust2}Dust aggregates are generated by impacting dust projectiles with velocities up to 7\,$\rm{ms^{-1}}$.}
\end{figure}
Targets were placed in a vacuum chamber at the bottom of a tube. After evacuating the whole tube, dust particles of different sizes were ejected through a sieve placed in the tube at different heights. Accelerated by gravity in free fall, dust aggregates impact the target with predefined velocities  between 1\,$\rm{ms^{-1}}$ and 7\,$\rm{ms^{-1}}$ (Fig. \ref{fig:uneffecteddust2}). 
In a first set of experiments, we extended the study of \citet{TeiserEtal2011b} to larger projectiles. \citet{TeiserEtal2011b} used a sieve with a mesh size of 125\,$\mu$m and produced dust particles with a mean size of 250\,$\mu$m. By using a sieve with a mesh size of 4\,mm, we were able to produce dust projectiles with a mean particle size of 1.2\,mm. This is an increase in the mass of the projectiles by about a factor of 100.

In a second set of experiments, we first produced highly porous targets through impacts at low velocities. We formed these highly porous targets by ejecting dust through the sieve of mesh size 125\,$\mu$m directly held above a transparent box at normal air pressure. 
We then raised the sieve to increase the impact velocity on the prepared highly porous target. 
In different experiment runs, we varied the sizes of the impacting fast dust-aggregates by using sieves with mesh sizes of 125\,$\mu$m and 4\,mm. We marked the boundary of the porous target with a colored layer of dust to illustrate the depth to which impacts influence the target below.

\section{\label{sec:resultsint}Results and interpretations}

In the following we present our results and provide some interpretations. 

\subsection{\label{sec:CM}Compression measurements}
\begin{figure}[h]
\includegraphics[width=\columnwidth]{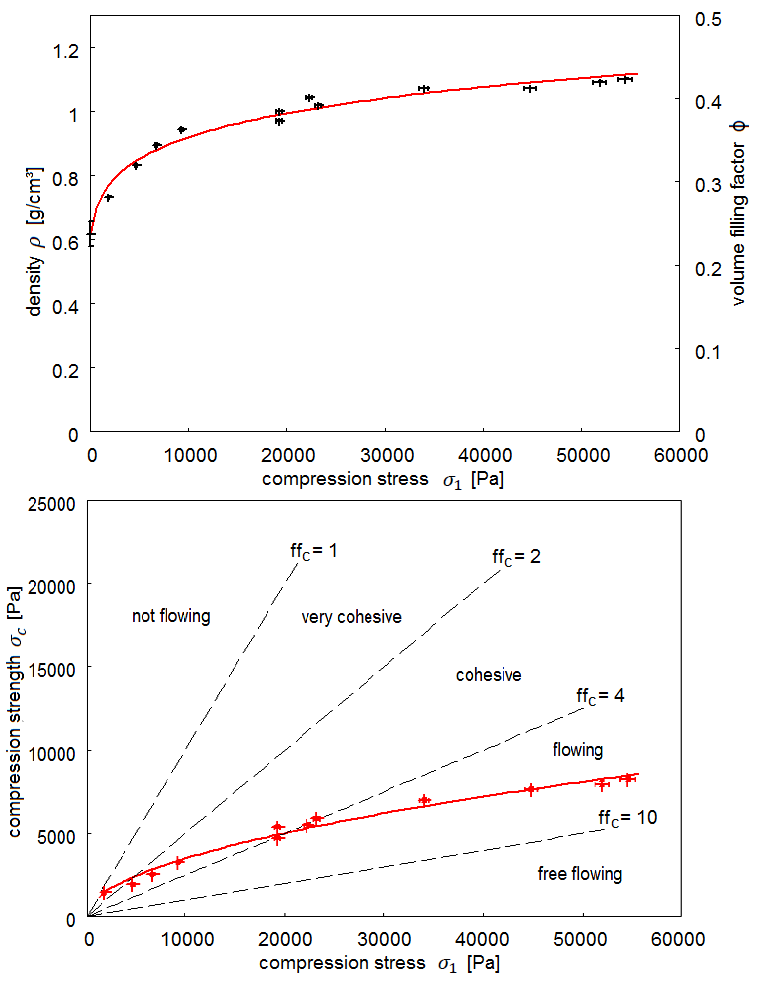}
    \caption{\label{fig:flowfunc}Top: density and filling-factor dependence of quartz dust on the compressive stress; Bottom: uniaxial compression strength as a function of the omnidirectional compression stress. Values for the flow function $\rm{\sigma_{c}}$ = f($\rm{\sigma_{1}}$) can be extracted from the intersections of the dashed lines with the data (red line).}
\end{figure}

\textbf{In the case of compression during sample preparation}, Fig. \ref{fig:flowfunc} (top) shows the density and volume-filling factor of the dust cylinders over compression stress applied to the confining mold
for production of the cylinders. 
By fitting a power law 
\begin{equation}
\rm{\rho} = \rho_{0} \cdot (1 + \frac{\rm{\sigma_{1}}}{\rm{\sigma_{0}}})^{n}
\end{equation}
characterizing the isentropic compression of bulk material, we can classify the $\rm{SiO}_{2}$ dust used by a 
calculated index of compressibility $ n $ \citep{Tomas2000, Tomas2001}. Additionally, we obtain the isostatic tensile strength $\rm{\sigma_{0}}$ and the initial density $\rm{\rho_{0}}$ of the dust placed in the mold. The following values were determined: 
\begin{table}[h]
\centering
\caption{Parameters for isentropic compression of quartz dust samples}
\label{tab:parameters}
\begin{tabular}{|c|c|c|}
\hline
$\rm{\rho_{0}}$ (g/cm$^3$)& $\rm{\sigma_{0}}$ (Pa)& $\rm{n} $\\ \hline
$0.612 \pm 0.024$ & $295.26 \pm 155.70$ & $0.115 \pm 0.009$\\ \hline
\end{tabular}
\end{table}
the index of compressibility characterizes the compressibility and cohesion / flow ability according to
the distinction of table \ref{tab:flowability1} \citep{Tomas2001}. In our case, the dust cylinders are classified as very compressible and very cohesive.
\begin{table}[h]
\centering
\caption{Semi-empirical classification of the index of compressibility}
\label{tab:flowability1}
\begin{tabular}{|c|c|c|}
\hline
index n & evaluation & flow ability\\ \hline
0 - 0.01 & incompressible & free flowing\\
0.01 - 0.05 & low compressibility & (free) flowing\\
0.05 - 0.1 & compressible & cohesive\\
0.1 - 1 & very compressible & very cohesive\\ \hline
\end{tabular}
\end{table}

\textbf{In the case of unidirectional compression}, an additional classification can be based on the flow function $\rm{ff_{c}}$, which is defined as the ratio of the compression stress $\rm{\sigma_{1}}$ to the compression strength of the samples $\rm{\sigma_{c}}$
\begin{equation}
\rm{ff_{c}} = \frac{\rm{\sigma_{1}}}{\rm{\sigma_{c}}}.
\end{equation}
The flow function is illustrated in Fig. \ref{fig:flowfunc} (bottom), which shows the compression strength versus compression stress.
The data show a typical trend, as more solidified dust cylinders have greater resistance to the applied unidirectional pressure.   
The flow function leads to the following classification as given in table \ref{tab:flowability2} \citep{Jenike1964}.  The characterization varies between  "not flowing" for $\rm{ff_{c}} < 1$ and "free flowing" for $\rm{ff_{c}} > 10$:
\begin{table}[h]
\centering
\caption{Flow ability of granular matter or powders \citep{Jenike1964}}
\label{tab:flowability2}
\begin{tabular}{|l|l|}
\hline
$\rm{ff_{c}} < 1$ & non flowing\\
$1 < \rm{ff_{c}} < 2$ & very cohesive\\
$2 < \rm{ff_{c}} < 4$ & cohesive\\
$4 < \rm{ff_{c}} < 10$ & easy flowing\\
$10 < \rm{ff_{c}}$ & free flowing\\
\hline
\end{tabular}
\end{table}
We find that low density cylinders (low applied compression stress) fall into a class of being cohesive or very cohesive in agreement with the classification given above. In contrast to that, the most compact dust-aggregates are characterized as flowing. This does not contradict the classification from the power-law fit for the isentropic compression given above, as this is a single fit to the whole range of compression stresses but is mostly sensitive to the low compression stress values where the density changes 
significantly. If the behavior of the dust aggregates were also divided between low and high compression stress,
the data for the high compression stress would be in agreement with a much lower index of compressibility and therefore flowing conditions (Fig. \ref{fig:flowfunc}, top). The flow function seems to be a more reliable tool here to characterize dust aggregates of
different filling-factors. 

The resulting classification then also agrees with general observations that aggregates 
of either low density or low filling factor can more
easily lose energy in collisions by restructuring the constituent grains and compaction when dense aggregates are more elastic \citep{BeitzEtal2011}.

\subsection{\label{sec:TSM}Tensile strength measurements}
\begin{figure}[h]
\includegraphics[width=\columnwidth]{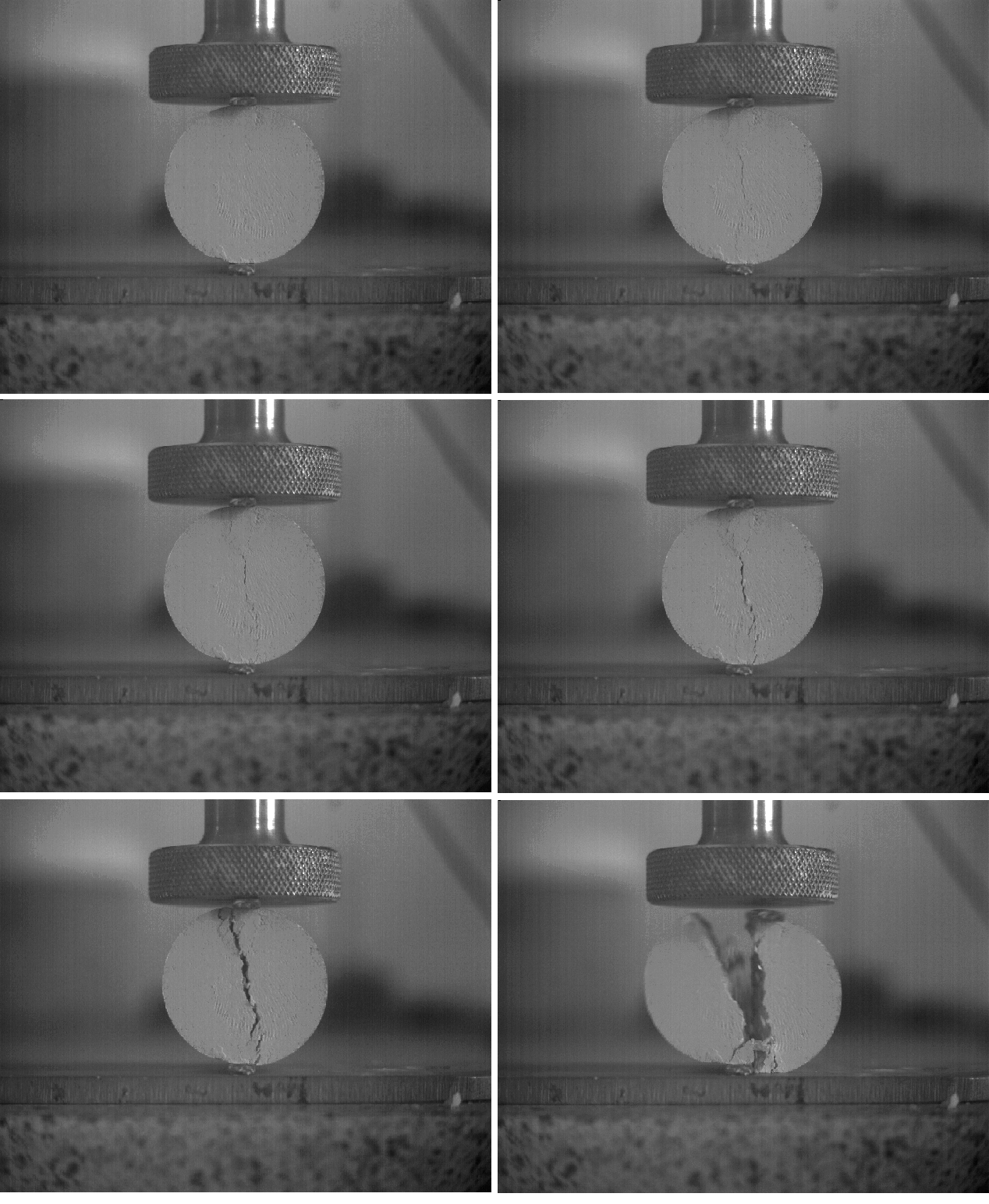}
    \caption{\label{fig:bratest1}Image sequence of the disruption of a dust cylinder into two halves during a Brazilian test. The expansion of the fracture from the middle towards the bars is clearly visible.}
\end{figure}

\begin{figure}[h]
\includegraphics[width=\columnwidth]{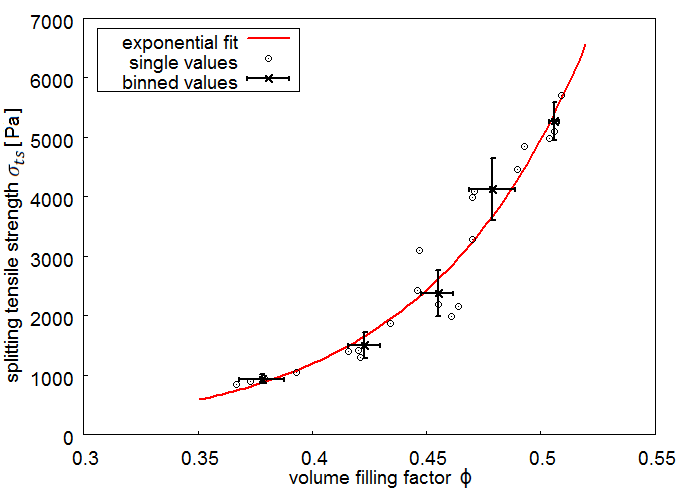}
    \caption{\label{fig:bratension1}Splitting tensile strength over density measured for 21 dust cylinders (open circles). The data were binned (black crosses). The red line shows an exponential fit.}
\end{figure}

During the Brazilian test shown in Fig. \ref{fig:brazil}, dust cylinders with different densities were observed to crack into two halves. An example image sequence is shown in Fig. \ref{fig:bratest1}.
Additionally, two very small wedged pieces of dust caused by the use of the load spreading bars, were produced at the break. They developed directly above and below the bars. According to \citet{Mitchell1961}, this is an indication of a shear fracture. We also observed that the fracture developed in the middle of the cross-section and propagated in the direction of the two bars. This is in agreement with the investigations of \citet{RoccoEtal1999}. 
The dependence of the splitting tensile strength on the volume filling-factor is seen in Fig. \ref{fig:bratension1}. 
With an increase in volume filling-factor $ \phi $, the dust aggregates show more resistance or have larger values of the splitting tensile strength $\rm{\sigma_{ts}}$. Without seeking further justification but describing the increase in the tensile strength in terms of increasing filling-factor, the most convenient fit turned out to be the exponential dependence
\begin{equation}
\rm{\sigma_{ts}} = \rm{4 \cdot e^{14.3 \cdot \phi}}
\end{equation}

\subsection{\label{sec:SVYM}Sound velocity measurements and modulus of elasticity (Young's modulus)}
\begin{figure}[h]
\includegraphics[width=\columnwidth]{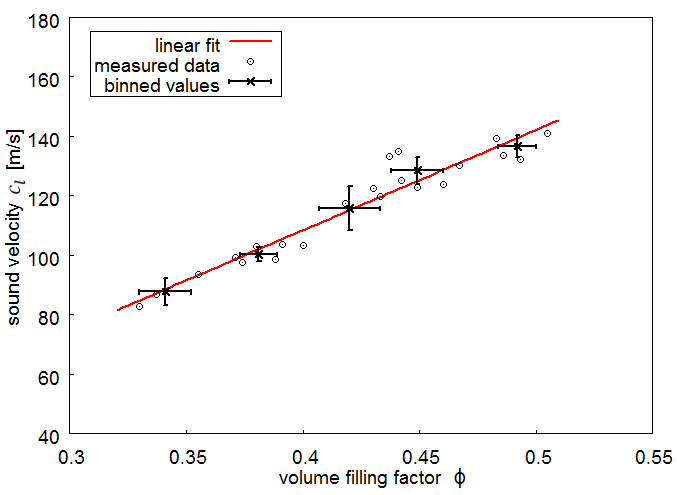}
    \caption{\label{fig:schallg}Measured data for sound velocity versus filling factors.}
\end{figure}

In Fig. \ref{fig:schallg}, we present our measured data for the sound velocity. With increasing filling-factors, a linear increase in the sound velocity $\rm{v_{s}}$ can be detected, which we fitted as
\begin{equation}
\rm{v_{s}} = \rm{337 \cdot \phi \, m/s - 26.4 \, m/s}.
\end{equation}
The absolute values are small in comparison with sound velocities in monolithic quartz ($v_{s} \approx 5000\,\rm{m/s}$). The obtained sound velocities in powdery $\rm{SiO}_{2}$ samples are several 10 times smaller. We note that the comparison to measurements not done 
in a vacuum but at an ambient pressure of 1 atmosphere showed no significant difference.

As the modulus of elasticity is not measured easily for porous dust-aggregates, the sound velocity provides a means for an indirect
determination. The sound velocity is
\begin{equation}
\rm{c_{l}} = \sqrt{\frac{\rm{E(1 - \nu)}}{\rm{\rho(1 + \nu)(1 - 2\nu)}}},
\end{equation}
where $ E $ is the modulus of elasticity, $\rm{\rho}$ is the density, and $\rm{\nu}$ is the lateral contraction or Poisson number. By neglecting the Poisson number, which is usually small ($\rm{\nu}$ $\approx$ 0), and solving the equation for \rm{E}, we obtained the estimation for our modulus of elasticity
\begin{equation}
E = \rm{c_{l}}^{2} \cdot \rho
\end{equation}

\begin{figure}[h]
\includegraphics[width=\columnwidth]{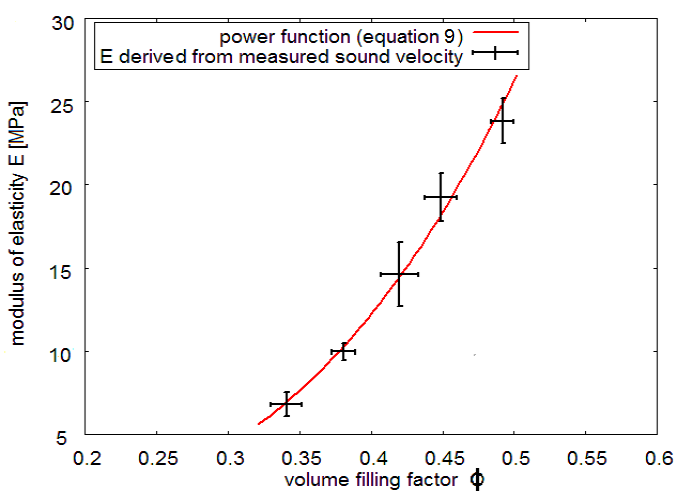}
    \caption{\label{fig:elastmodul}Calculated moduli of elasticity based on the measured sound velocities.}
\end{figure}

To calculate the moduli of elasticity, we performed a linear fit for the sound velocity and identified $\rm{v_{s}}$ with $\rm{c_{l}}$. Together with \mbox{$\rho = 2600\,\rm{kg/{m}^{3}} \cdot \phi$}, we derive
\begin{equation}
E = 2.96 \cdot 10^8 \cdot \phi^3 - 4.68 \cdot 10^7 \cdot \phi^2 + 1.82 \cdot 10^6 \cdot \phi \cdot  \left( \rm{\frac{N}{m^2}}  \right).
\end{equation}
The resulting moduli of elasticity are shown in Fig. \ref{fig:elastmodul}. 
The obtained data are very low values for the moduli of elasticity or Young's modulus (on the order of $10\,$ MPa). The modulus of elasticity for solid $\rm{SiO}_{2}$ is given by  $85 $ GPa \citep{Carlotti1995} which is about three orders of magnitude larger.

\subsection{\label{sec:DTS}Results and observations of collision experiments with dust particles}   
   
\subsubsection{\label{sec:PDG}Homogeneous aggregate formation}
\begin{figure}[h]
\includegraphics[width=\columnwidth]{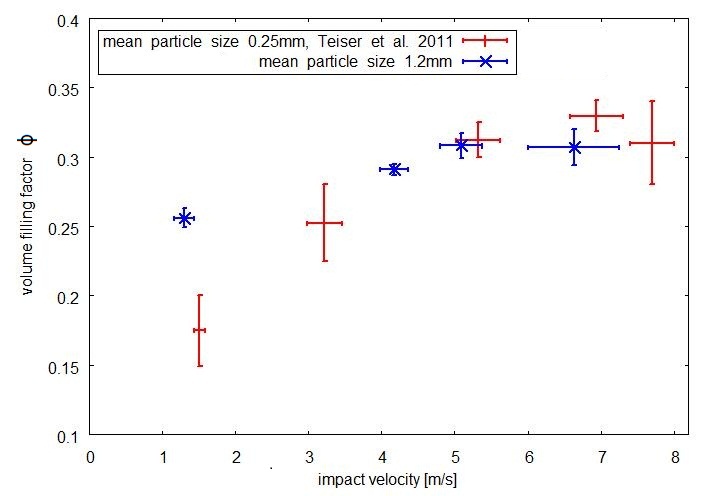}
    \caption{\label{fig:volumefilling}Volume filling-factors of dust aggregates generated by free falling impacting dust aggregates
    for two different projectile sizes: blue 1.2\,mm, this study;  red 0.25\,mm, from \citet{TeiserEtal2011b}}
\end{figure}

\citet{TeiserEtal2011b} used particle sizes of 0.25\,mm to grow dust aggregates. In this study, we used a larger mesh size for the sieve. This provides dust particles with mean particle sizes of 1.2\,mm. Approximating the mass by the third power of the size, this is a factor of 100 more in mass. The dependence of the 
filling-factor on the impact velocity is compared in Fig. \ref{fig:volumefilling} with the data measured by \citet{TeiserEtal2011b}.
At lower velocities (approximately up to 3\,m/s), the generated dust aggregates are more compact if built from larger particles. This is understandable as they possess more energy and can restructure the target surface more efficiently. However, once the collision velocity 
reaches 5\,m/s, the forming aggregates reach the same limiting volume filling-factor of $0.31 \pm 0.01$.  

\subsubsection{\label{sec:LDA}Layered aggregates}
\begin{figure}[h]
\includegraphics[width=\columnwidth]{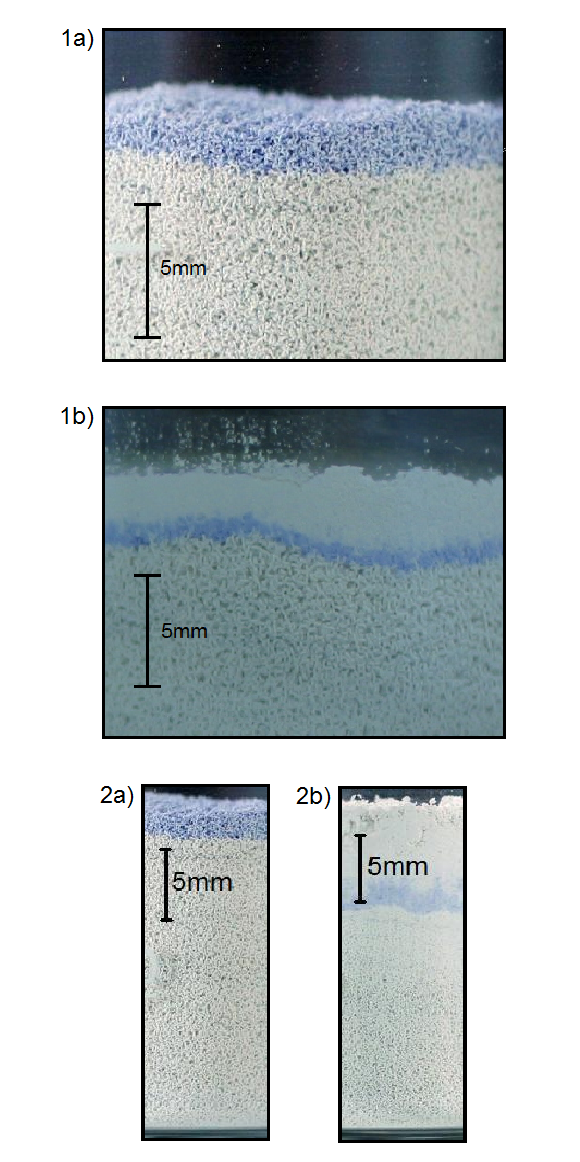}
    \caption{\label{fig:layers5}Dependence of the volume filling factor on impacting projectiles; 1a, 2a: low filling factor aggregates $\phi=0.14$ generated by sub m/s impacts; 1b: change of original target upon collisions with 0.25 mm aggregates at 6 m/s; 2b: change of original target upon collisions with 1.2 mm aggregates; The transition zone is only on the order of up to several mm. It separates two protoplanetesimal zones with filling factors characterized by the process by which they formed, i.e. a core-shell particle with highly
    porous core and dense shell forms.}
\end{figure}

As basic targets in this set of experiments, we used highly porous dust targets formed by slow impacts of aggregates produced of 0.25\,mm aggregates  (Fig. \ref{fig:layers5} top). These highly porous targets still show the characteristic size of the projectiles in their texture.
The initial aggregates (Fig. \ref{fig:layers5}, 1a and 2a) have a mean volume filling-factor of 0.14. We then exposed these targets to impacts at higher velocities of about 6\,m/s. As
aggregate size, we used in the first case (Fig. \ref{fig:layers5}, 1b) the small 0.25\,mm and in the second case the large 1.2\,mm particles (Fig. \ref{fig:layers5}, 2b). 
The small projectiles build up a compact layer above the initial target sample. The initial target is unchanged below the marked layer. Only the initial, about 1\,mm thick, marked layer of the target is therefore influenced by the impacts (Fig. \ref{fig:layers5}, 1b). A more extended compression zone of about 6.5\,mm was observed for the larger projectiles (Fig. \ref{fig:layers5}, 2b), even thoughthe initial target remained unchanged below. A quantification of the 
transition layer depending on velocity is currently beyond the capabilities of the setup. However, the experiments show that a highly porous dust- aggregate of cm size or larger can keep this porosity in its core if collision velocities increase and a more compact shell is added. Core mantle aggregates would be the result with inhomogeneous filling-factors.

\section{\label{sec:discussion}Application and conclusions}

The collisional evolution from $\rm \mu m$-sized dust grains to km-sized planetesimals and beyond strongly depends on the mechanical properties of dust aggregates. The basic mechanical properties are the modulus of elasticity, the tensile strength, and the compression strength. 
These parameters are intrinsically difficult to determine for dust aggregates. It is difficult to measure the modulus of
elasticity directly, as a small amount of stress compresses the aggregates, changing both the porosity and modulus of elasticity. 
However, by measuring the speed of sound, which is related to the Young's modulus,  
we were able to determine the porosity dependence of a dust sample. 
We also succeeded in applying the Brazilian test to dust aggregates of moderate density and determining the tensile strength of the same kind of dust aggregate. Determining the compression strength in addition gives a set of properties, which allows us to develop a more realistic model for dust aggregates in numerical simulations of collisions, though the set
 is still incomplete as e.g. shear is not considered. 
In total, different conclusions can be drawn from our experiments, which are listed first and detailed afterwards. 

\begin{itemize}

\item Most basicly, the results can be used as input parameters to detailed numerical simulations of collisions. In this respect, the measurements allow no further conclusion here. Implications have to await their use in such simulations.

\item The porosity evolution in a self-consistent growth scenario can be compared to earlier speculations by collecting small dust aggregates at a given velocity.

\item The porosity is important as it strongly influences the thermal properties of dust aggregates and our results can be used to quantify possible consequences, e.g. for photophoretic particle sorting in protoplanetary disks.

\item Even without detailed simulations, the outcome of published collision experiments between cm-size aggregates can be 
re-interpreted in the light of the determined properties, depending on possible porosity evolutions.

\item Sound propagation in cohesive and granular matter are part of the research in communities apart from planet formation and our measurements can be compared to more fundamental models. The exact value is also important in helping us to decide whether
a collision is supersonic or subsonic, which has an impact on the way in which we model dust-aggregate collisions.

\end{itemize}
As noted, not much more can be said about the first item, which is an input parameter to numerical simulations. It is known that variations in material properties can strongly influence the outcome of a collision. Depending on either the porosity or the existence of a
less porous (hard) shell on a porous core, two aggregates might stick, bounce off each other, or fragment \citep{GeretshauserEtal2011b, GeretshauserEtal2010, GeretshauserEtal2011}. The full importance of our measured data in this range of outcomes can only be seen in detailed simulations that remain to be carried out in the future.
Typical values of the parameters studied are in the following range:

\begin{itemize}
\item compression strength (unidirectional) 1 -- 8 kPa\\
(for omnidirectional compression and flow function see section \ref{sec:UCT}),
\item tensile strength 1 -- 5 kPa,
\item modulus of elasticity 5 -- 25 MPa,
\item sound speed 80 -- 140 m/s.
\end{itemize}

Nevertheless, the measured properties already shed some light on recent collision experiments. 
\citet{BeitzEtal2011} studied the collisions between cm-size dust aggregates,
which had porosities in the same range as the porosities studied here and the same kind of dust (micron-sized quartz). They found that in a collision at about 1\,$\rm{ms^{-1}}$, a slightly more porous aggregate gets destroyed and adds some mass to the more compact aggregate, which stays intact even if the porosity difference is only 0.02. This can be qualitatively explained by having a look at the tensile strength (Fig. \ref{fig:bratension1}), which shows an exponential dependence on the volume filling-factor. Therefore, a slightly more compact aggregate is much more stable. It was not expected that this effect be so strong nor that very small variations in porosity could decide whether an aggregate fragments or growth. This clearly shows that the porosity evolution during planetesimal formation is a key factor in the process of planetesimal formation.

While they were meant to be indirect tools for determining the modulus of elasticity, the speed of sound measurements are also important in their own respect. They vary between 80\,$\rm{ms^{-1}}$ and 140\,$\rm{ms^{-1}}$ across the range of studied porosities. This is much lower than the speed of sound in monolithic solid material. Numerical simulations by \citet{PaszunDominik2008} and \citet{RinglUrbassek2012} are consistent to within an  order of magnitude, considering the different properties of the underlying dust grains. 
For moderate models of protoplanetary disks, the absolute values do not differ considerably but are clearly above the collision velocities of up to 50\,$\rm{ms^{-1}}$ \citep{WeidenschillingCuzzi1993}. However, somewhat more turbulent disk collisions might become supersonic rather early, which can influence any subsequent collisions.

A few more aspects of collisional evolution come from our collision experiments in this work. 
They show that the average properties of a whole aggregate are appropriate if this aggregate grows by a continuous process,  i.e. by adding small aggregates at a given collision velocity. 
The volume filling factor depends on the collision-velocity but we note that has a limit of about 0.31 at 
high collision velocities. We extended earlier experiments to larger projectile sizes here. The experiments show that the limiting filling-factor is independent of the projectile mass when it is increased by a factor of 100. Using the same dust material, \citet{Kothe2010} found much larger filling factors in multiple-collision experiments. \citet{Kothe2010} found a filling factor of 0.4 at a velocity of 6\,$\rm{ms^{-1}}$ and extrapolated their results to values of even 0.6 at 10\,$\rm{ms^{-1}}$. 

As discussed before, the filling factor of a growing aggregate is very important, and the results of \citet{Kothe2010} and \citet{TeiserEtal2011b} may indicate quite different evolution 
during further growth. It therefore remains unclear how we can resolve this significant discrepancy, which is one reason why we extended the collisions
to larger projectile sizes, close to the size used by \citet{Kothe2010}. After we had used the same projectile masses, only one difference remained in the experiments. The targets of \citet{TeiserEtal2011b} and this work were built by multiple impacts at random sites of a target surface, while \citet{Kothe2010} studied collisions of always the same site with a new aggregate. This created a tip of dust that is smaller than the next projectile and is repeatedly compressed. Surface structures in our experiments are small compared to the projectile size, and the local pressure of an impacting projectile is spread over a larger surface area. We propose that
this difference in setup as an explanation of this discrepancy. Both situations might occur in protoplanetary disks. If very irregular aggregates with sharp extensions (i.e. from earlier catastrophic disruptions) were hit by the next projectile, local
filling-factors might be as high as determined by \citet{Kothe2010}. However, if 
random impacts created rather smooth surfaces, i.e. a growing aggregate moving through a cloud of small aggregates, 
a lower filling factor with a limiting value of about $\Phi = 0.31$ would result.

Following up on this are the experiments where aggregates were launched at higher velocities up to 6\,$\rm{ms^{-1}}$ but onto a highly porous dust target formed in low speed collisions. The experiments show that the existing part of the aggregate is rather insensitive to these impacts. Only the new layer that builds up consists of dust with a filling factor characteristic of the impact speed, i.e. that is compressed to a maximum filling-factor of 0.31. The original aggregate can keep its original high porosity. In this way, as impact speed increases with growing target size in protoplanetary disks, dust aggregates may consist of a porous core and a compact shell. This influences the particle-gas coupling time, which determines the collision velocities of subsequent collisions, and will likely be important for collisional outcomes at larger impact speeds at later phases where collisional effects reach the core. \citet{GeretshauserEtal2011b} showed that the mechanical properties of these shell-core-aggregates change with the ratio of core to shell mass, but is almost dominated by the properties of the outer shell.

\citet{krause2011} determined the thermal conductivity of dust aggregates in experiments. They varied the filling factor and found that more porous dust-aggregates have lower thermal conductivities.
In addition, they found mm-size particles drift through a protoplanetary disk at low speed $(\approx m/s)$ but that this speed still varies significantly across this size range. Our experiments show that the filling factor of growing aggregates is still below the limiting filling-factor of 0.31. As it increases with velocity and as larger particles are faster, the filling factor increases with the size of the aggregates. 
This, means in turn, that larger aggregates have higher thermal conductivity.
\citet{LoescheWurm2012} used these results to estimate the photophoretic strength of dust-mantled chondrules, and showed that this evolution of filling factors allows a sorting of the particles according to size as found
in primitive meteorites \citep{kuebler1999}. Our experiments show that this is a realistic estimate.

The experiments and their applications to planet formation show that the filling factor of dust aggregates 
is  one of the most important parameters determining the physical evolution of small bodies in protoplanetary disks.
We found that smaller (cm-size) aggregates are unlikely to be compact as they could be, but eventually reach a limiting filling-factor of 0.31 and have a core-shell structure. Since the more compact aggregates are more robust in collisions,
they can grow further by feeding on the less compact aggregates. In earlier experiments, \citet{wurm2005b} and \citet{TeiserEtal2011a} demonstrated that larger aggregates with a filling factor of 0.31 can grow even by the impact of high speed projectiles. 
\citet{WindmarkEtal2012} showed that this process allows the collisional growth of planetesimals. These earlier collision experiments used
an arbitrary filling-factor of 0.31 for the manually prepared targets and projectiles. We have found and confirmed that this choice was well made as the limiting filling-factor found in our experiments of self-consistent impact growth is exactly the same filling-factor as used in those earlier experiments. We conclude that this is strong support for a collisional growth scenario of planetesimals.

\begin{acknowledgements}
      This work is funded by the 
      \emph{Deut\-sche For\-schungs\-ge\-mein\-schaft, DFG\/ as part of the research group FOR 759}. We also
      thank the referee for a very constructive review of our manuscript.
\end{acknowledgements}

\bibliographystyle{aa} % style aa.bst
\bibliography{protoplanetesimals} % your references Yourfile.bib

\end{document}